\documentclass[12pt]{iopart}
\usepackage{graphicx}
\usepackage{amssymb}
\usepackage[dvips]{color}
\newcommand{\bfr}{{\bf r}}

\newcommand{\bfq}{{\bf q}}

\newcommand{\um}{{U_m}}
\newcommand{\hpsi}{{\hat\psi}}
\newcommand{\hPsi}{{\hat\Psi}}
\newcommand{\hphi}{{\hat\phi}}
\newcommand{\dpsi}{{\hat{\delta\psi}}}
\newcommand{\dPsi}{{\hat{\delta\Psi}}}
\newcommand{\ha}{{\hat a}}
\newcommand{\hb}{{\hat b}}
\newcommand{\halpha}{{\hat \alpha}}
\newcommand{\omz}{{\omega_0}}
\newcommand{\lz}{{l_0}}
\newcommand{\lm}{{l_{AB}}}
\newcommand{\labc}{{l_{ABC}}}

\newcommand{\nav}{{\bar n}}

\begin{document}
\title{Confinement controlled dissociation of a molecular Bose-Einstein condensate}
\author{I. Tikhonenkov and A. Vardi}
\address{Department of Chemistry, Ben-Gurion University of the
Negev, P.O.B. 653, Beer-Sheva 84105, Israel}
\ead{avardi@bgu.ac.il}

\begin{abstract}
We study the collective two-channel dissociation dynamics of a molecular Bose-Einstein condensate into bosonic fragments under tight harmonic confinement. Bose-stimulated dissociation in either channel can only take place provided that the respective trap size $l_i$ for the fragments is large with respect to the healing length $\zeta_i$ of the atom-molecule resonance. Thus, even when both channels are equally coupled, differences in mass or in dynamical polarizability enable the control of the reaction outcome by variation of the trap frequency. In particular, if $l_1>\zeta_1$ and $l_2<\zeta_2$, only the first channel will be amplified. This behavior demonstrate a unique feature of 'superchemistry' wherein a chemical reaction may be controlled by the manipulation of the reaction vessel.
\end{abstract}   

\submitto{\JPB}

\pacs{03.75.Kk 03.75.Mn 42.50.Fx}

\section{Introduction}

Quantum degenerate bose gases of dimers consisting both fermionic \cite{Regal03,Strecker03,Cubizolles03,Jochim03,Zwierlein03} and bosonic \cite{Herbig03,Durr04,Xu03} constituent atoms, have recently been realized in Feshbach resonance \cite{Tiesinga93,Inouye98,Stenger99,Timmermans99} and photoassociation \cite{Band95,Fedichev96,Vardi97,Julienne98,Wynar00,Fatemi00,Rom04,Winkler05,Ryu06,Stoferle06} setups. First evidence for boson pairing was obtained in seminal experiments in a BEC of $^{85}$Rb atoms \cite{Donley02} showing Ramsey-like oscillations indicative of the coherent coexistence of molecular and atomic condensates. This experiment was soon followed by the production of much larger and well-defined molecular condensates of $^{133}$Cs$_2$ \cite{Herbig03}, $^{87}$Rb$_2$ \cite{Durr04}, and $^{23}$Na$_2$ \cite{Xu03}, as well as ultracold molecular gases and molecular BECs with fermionic constituent atoms, such as $^{40}$K$_2$ \cite{Regal03} and $^6$Li$_2$ \cite{Strecker03,Cubizolles03,Jochim03,Zwierlein03}. BECs of molecules consisting fermionic atoms,turn out to be more stable due to the Pauli blocking of collisional relaxation \cite{Petrov04}. 

These novel entities open the way for the corroboration of theoretical proposals for,  so called, superchemistry \cite{Javanainen99,Heinzen00,Vardi02,Moore02,Olsen04,Jack05}. The underlying dynamics of atom-molecule conversion in quantum gases, is collective in essence and is therefore strongly affected by the quantum statistics of atoms and molecules. For the purely bosonic system, bose enhancement takes place for atoms and molecules, to result in a nearly exact analogy \cite{Javanainen99,Drummond98,Vardi01,Meystre05,Rowen05} with parametric processes in optics, including coherent matter wave amplification, phase conjugation, and quantum state squeezing. Association of atoms to form a condensate of dimers is analogous to second-harmonic generation and the dissociation of a molecular condensate corresponds to parametric downconversion. The latter pair-production process is known to exhibit a modulational instability, resulting in from the exponential amplification of spontaneously emitted photon-pairs \cite{Yariv,Walls,Shen,Scully}. 

The existence of similar parametric gain in molecular dissociation was noted early on in the theoretical study of atom-molecule condensates \cite{Javanainen99,Vardi01,Vardi02,Moore02,Jack05,Heinzen00}. Moreover, since driving the system in the vicinity of a dynamical instability generates gain in quantum noise \cite{VardiAnglin01}, there are significant deviations from the classical-field predictions of the Gross-Pitaevskii (GP) mean-field theory \cite{Vardi01,Hope01,Poulsen01,Jin05,Naidon06}. The rapid growth of correlations can be used for the generation of pair-correlated or number-squeezed atomic beams \cite{Kheruntsyan02,Kheruntsyan05} using either spontaneously emitted pairs \cite{Kheruntsyan02} or a classical atomic (signal) seed \cite{Kheruntsyan05}, in analogy to the use of parametric downconversion to generate squeezed light in quantum optics \cite{WallsYurke}.

Recently, we have found that stimulated molecular dissociation can be arrested, provided that atoms are confined to move within an harmonic trap of frequency $\omz$, whose size $\lz=\sqrt{\hbar/(m\omz)}$ is smaller than the resonance healing length $\zeta=\hbar(mg)^{-1/2}\nav^{-1/4}$ \cite{Tikhonenkov07}, where $m$,$g$, and $\nav$ are the atomic mass, the atom-molecule coupling strength, and the average molecular density, respectively. Thus, the large coherence length of the molecular BEC on the one hand, and the tight confinement available today by magnetic and optical traps on the other hand, conjure to produce collective chemistry which not only depends on particle statistics, but also on the size and shape of the 'reaction vessel' itself. Here we show that for two-channel dissociation of a molecular condensate of triatomic $ABC$ molecules, the critical dependence on the trapping potential can be used to control the dissociation outcome by mere manipulation of the trap parameters, even when the two rearrangement channels are equally coupled to the molecular BEC. Thus for sufficiently tight traps, neither channel is amplified. At the other extreme, for trap size larger than the resonance healing-length of both channels, both channels will be amplified. In between these regimes, lies a window where only one channel attains exponential gain, resulting in nearly perfect selectivity.

The exponential gain features of molecular BEC dissociation and the possibility to arrest the dissociation of a homonuclear diatomic quantum gas by tight confinement are presented in Section \ref{homonuc}. The extension to the case of heteronuclear molecules where dissociation fragments have different masses and optical polarizabilities, resulting in differences in trap-size and healing length, is provided in Section \ref{heteronuc}. In Section \ref{multi} we present the analysis for the two-channel dissociation of a triatomic quantum gas into bosonic constituents and corroborate the analytical prediction by numerical examples. The values of experimental parameters required for the observation of the predicted phenomena, are briefly discussed in Section \ref{experiment}, while conclusions and future directions are presented in Section \ref{conclusion}

\section{Exponential gain and stabilization by confinement} 
\label{homonuc}

In this section we provide a brief review of the results of Ref. \cite{Tikhonenkov07} which serve as the foundation for confinement control.  We consider the dissociation of a molecular BEC constituting dimers of identical bosonic atoms. The atom-molecule Hamiltonian in the rotating wave approximation is,
\begin{eqnarray}
\label{ham}
H&=&\int d\bfr\left\{\hpsi^\dag\left[-\frac{\hbar^2}{2m}\nabla^2+V(\bfr)+\Delta\right]\hpsi\right.\\
~&~&\left.+\hphi^\dag\left[-\frac{\hbar^2}{4m}\nabla^2+2V(\bfr)+\frac{\um}{2}\hphi^\dag\hphi\right]\hphi+\left[\frac{g}{2}\hpsi^\dag\hpsi^\dag\hphi+H.c.\right]\right\}~,\nonumber
\end{eqnarray}
where $\hpsi(\bfr,t)$ and $\hphi(\bfr,t)$ are atomic and molecular bose field operators, $m$ is the atomic mass, $\Delta$ is the detuning from atom-molecule resonance (i.e. the interaction-free energy difference between the threshold of the open channel and the molecular bound state in the closed channel of a Feshbach-coupled system, or the detuning from two-photon resonance in a photodissociation scheme) and $V(\bfr)=m\omz^2 r^2/2$ is an isotropic optical trap potential with frequency $\omz$. Atoms and molecules see essentially the same trap frequency since molecules have twice the mass and twice the optical polarizability \cite{Greiner03}. The molecule-molecule interaction strength is $\um$ and $g$ is the atom-molecule Feshbach or optical coupling frequency, assumed without loss of generality to be real. Interactions between atoms are neglected since we are interested in the initial dissociation dynamics when atomic densities are small.

The exponential gain features of molecular BEC dissociation are readily captured within an undepleted pump approximation. This amounts to linearization of the atom-molecule Hamiltonian (\ref{ham}) about a constant molecular BEC wavefunction $\hphi(\bfr,t)\rightarrow\phi(\bfr)e^{-i\mu t/\hbar}$, obeying the Gross-Pitaevskii time-independent equation 
\begin{equation}
-\frac{\hbar^2}{4m}\nabla^2\phi+[2V(\bfr)+\um|\phi|^2]\phi=\mu\phi~.
\label{mgp}
\end{equation} 
The dynamical equations for atomic field operators $\hPsi=\hpsi e^{i\mu t/ 2\hbar}$, and $\hPsi^\dag=\hpsi^\dag e^{-i\mu t/ 2\hbar}$, read,  
\begin{eqnarray}
\label{stabo}
i\hbar\frac{\partial}{\partial t}\hPsi&=&\left[-\frac{\hbar^2}{2m}\nabla^2+\delta+V(\bfr)\right]\hPsi+g\phi\hPsi^\dag,\\
\label{stabt}
-i\hbar\frac{\partial}{\partial t}\hPsi^\dag&=&\left[-\frac{\hbar^2}{2m}\nabla^2+\delta+V(\bfr)\right]\hPsi^\dag+g\phi^*\hPsi,
\end{eqnarray}
where  $\delta=\Delta-\mu/2$ is an effective detuning including molecular zero point kinetic energy and molecular mean-field shift. 

This linearization procedure (which is used extensively in three- and four-wave mixing in optics \cite{Yariv,Walls,Shen,Scully}) is inherently limited to short-time dynamics where molecular condensate depletion as well as fluctuations of the molecular field are small. Its validity regime is set by the standard parametric-approximation limits: $gt\rightarrow 0$, $\phi\rightarrow\infty$ while $g\phi t=constant$ \cite{Scully}. Within these limits, it constitutes a reliable stability analysis towards stimulated dissociation. Molecular condensate stability is determined by the characteristic eigenvalues $\lambda$ of the set (\ref{stabo})-(\ref{stabt}). For a uniform gas with $V(\bfr)=0$, $\phi=\sqrt{n}$, and $\mu=\um n$ with $n$ being the molecular condensate density, the system is readily diagonalized in momentum space $\hpsi(\bfr,t)=\sum_\bfq \ha_\bfq(t)\exp(i\bfq\cdot\bfr)$, to give the dispersion relation,
\begin{equation}
\hbar\lambda_q=\sqrt{(\epsilon_q+\delta)^2-n|g|^2}~,
\label{eigen}
\end{equation} 
where $\epsilon_\bfq=(\hbar q)^2/2m$ is the free-particle dispersion. Thus for a uniform gas with $\delta<\sqrt{n}|g|$, there are always unstable resonant modes for which $\epsilon_\bfq+\delta<g\sqrt{n}$, resulting in complex characteristic frequencies and exponential gain. The dynamics of amplified modes is described by well-known solutions, given in terms of the hyperbolic functions $\cosh(\lambda_q t)$ and $\sinh(\lambda_q t)$, which can be found in Refs. \cite{Vardi01,Kheruntsyan02,Kheruntsyan05,WallsYurke}. For zero effective detuning $\delta=0$, the unstable modes are simply low-energy, long-wavelength excitations.

The stability analysis for a nonuniform molecular gas can be gleaned from the Bogoliubov stability analysis of an attractively interacting atomic BEC with the Hamiltonian,
\begin{equation}
\label{hamt}
H=\int  \hpsi^\dag\left[-\frac{\hbar^2}{2m}\nabla^2+V(\bfr)+\frac{U}{2}\hpsi^\dag\hpsi\right]\hpsi d\bfr~.
\end{equation}
The Bogoliubov equations are obtained from the usual separation $\hpsi(\bfr,t)=\psi(\bfr)+\dpsi(\bfr,t)$, where $\psi(\bfr)$ is the $c$-number solution of the time-independent GP equation $[-\frac{\hbar^2}{2m}\nabla^2+V(\bfr)+U|\psi|^2]\psi=\mu\psi$ and $\dpsi(\bfr,t)$ is an operator correction. Linearization, obtained by retaining  up to quadratic terms in $\dpsi$, $\dpsi^\dag$ and rotating $\dPsi=\dpsi e^{i\mu t/\hbar}$, $\dPsi^\dag=\dpsi^\dag e^{-i\mu t/ \hbar}$, gives the familiar forms,
\begin{eqnarray}
\label{bogo}
i\hbar\frac{\partial}{\partial t}\dPsi&=&\left[-\frac{\hbar^2}{2m}\nabla^2-\mu+V(\bfr)+2U|\psi|^2\right]\dPsi+U\psi^2\dPsi^\dag,\\
\label{bogt}
-i\hbar\frac{\partial}{\partial t}\dPsi^\dag&=&\left[-\frac{\hbar^2}{2m}\nabla^2-\mu+V(\bfr)+2U|\psi|^2\right]\dPsi^\dag+U{\psi^*}^2\dPsi.
\end{eqnarray}
Equations (\ref{bogo})-(\ref{bogt}) are similar to Eqs. (\ref{stabo})-(\ref{stabt}), except for the replacement of the classical molecular pump $\phi$ by the atom density $\psi^2$ and the addition of the diagonal mean-field term $2U|\psi|^2$ corresponding to normal pairing of the $\hpsi^\dag\hpsi^\dag\hpsi\hpsi$ interaction, which does not exist for the Feshbach coupling $\hphi^\dag\hpsi\hpsi+H.c$. The Bogoliubov transformation $\dPsi(\bfr,t)=\sum_\bfq \left(\halpha_\bfq u_\bfq(\bfr)e^{-i\lambda_\bfq t}-\halpha_\bfq^\dag v_\bfq^*(\bfr)e^{i\lambda_\bfq t}\right)$ seeks to diagonalize the set (\ref{bogo})-(\ref{bogt}) similarly to the diagonalization of Eqs. (\ref{stabo})-(\ref{stabt}). For a uniform bose gas in 3D, the additional term results in the modification of the dispersion relation to the famous Bogoliubov dispersion 
\begin{equation}
\hbar\lambda_\bfq=\sqrt{(\epsilon_q+nU)^2-(nU)^2}=\sqrt{\epsilon_q(\epsilon_q+2nU)}.
\label{bogdisp}
\end{equation} 
For an attractively interacting gas with $U=4\pi\hbar^2 a_s/m<0$ ($a_s$ being the $s$-wave scattering length of the atoms), long wavelength excitations with $\epsilon_q<|2nU|$ corresponding to phonon-like excitations for equal-magnitude but positive $U$, attain complex frequencies and are thus unstable, leading to the collapse of the condensate \cite{Stoof94,Gerton00,Donley01}. However, for a BEC subject to a confining potential, collapse can be arrested provided that the size of the trap $\lz$ is short with respect to the healing length $\xi$, i.e.
\begin{equation}
\lz\equiv\sqrt{\frac{\hbar}{m\omz}}<\sqrt{\frac{\hbar^2}{2m\nav|U|}}\equiv\xi,
\end{equation}
where $\nav$ denotes the average condensate density. Thus, provided that the spacing between trap levels $\hbar\omz$ is larger than the average interaction energy $|2\nav U|$, the unstable modes are precluded from the trap and stable condensates of attractively interacting atoms exist up to a critical density for a given trap-frequency \cite{Ruprecht95,Bradley}. The maximum number of particles allowed for a given trap, is obtained from writing the trap frequency and the interaction energy in terms of the trap size $\lz$,
\begin{equation}
\hbar\omz=\frac{\hbar^2}{m\lz^2}~~,~~U\nav=U\frac{N}{V}\approx U\frac{N}{(4/3)\pi\left(\sqrt{2}\lz\right)^3},
\end{equation}
where $V$ denotes the trap volume, corresponding to the volume of a 3D spherical box whose radius is equal to the characteristic size of the interaction-free ground state $\sqrt{2}\lz$. This gives the familiar stability condition
\begin{equation}
N<N_C\approx\frac{4\pi 2^{3/2}}{3}\frac{\hbar^2\lz}{2m|U|}=\frac{\sqrt{2}}{3}\frac{\lz}{|a_s|}~,
\end{equation}
stating that the critical number of particles in the attractively interacting BEC corresponds to the number of $s$-wave scattering lengths contained within the characteristic trap size. For $^7$Li in harmonic traps of order 100Hz, $\lz$ is of the order of a few $\mu$m, compared with the $\sim 30$ bohr scattering length, giving critical numbers $N_C$ of several hundreds to few thousands particles \cite{Bradley}. Interestingly, experimental measurements give $N_C|a_s|/\lz=0.46\pm 0.06$ \cite{Roberts98}. This value is in an excellent agreement with the prediction above of $\sqrt{2}/3=0.47$, despite the fact that the theoretical value is based on a crude approximation for the density. However, due to the rather arbitrary choice of trap radius (taken here to be the point at which the ground-state order parameter drops to $1/e$ of its maximal volume), the agreement appears to be incidental rather than profound.

Returning to molecular dissociation, a similar threshold for instability is obtained, despite the differences between the gapless Bogoliubov dispersion relation (\ref{bogdisp}) and the 'imaginary-gapped' molecular dissociation spectrum (\ref{eigen}). Provided that the trap size is smaller than the 'resonance healing length' $\zeta$, i.e.
\begin{equation}
\lz\equiv\sqrt{\frac{\hbar}{m\omz}}<\sqrt{\frac{\hbar^2}{m g\nav^{1/2}}}\equiv\zeta,
\end{equation}
or alternatively if $\hbar\omz>g\sqrt{\nav}$, there will be no unstable modes towards dissociation, even when $\delta=0$ (i.e. on-resonance coupling). The resonance healing length is the characteristic lengthscale of Eqs. (\ref{stabo}) and (\ref{stabt}), resulting in from the interplay of kinetic and resonance energies. Thus, when this lengthscale exceeds the system size imposed by tight confinement, the difference in zero-point motion between bound molecules and pairs of dissociating atoms, inhibits the dissociation process. This effect is a characteristic of 'Superchemistry' \cite{Heinzen00} in that the outcome of a coherent chemical process depends due to its collective nature, not only on the statistics of the constituent atoms, but also on the size and geometry of the 'container' in which it is carried out. It results in from the fact that the entire gas behaves essentially as a macroscopic single-particle with a relatively large coherence length, and from the possibility to introduce confining potentials on the same $\mu$m lengthscale.
 
\begin{figure}
\centering
\includegraphics[scale=0.7,angle=0]{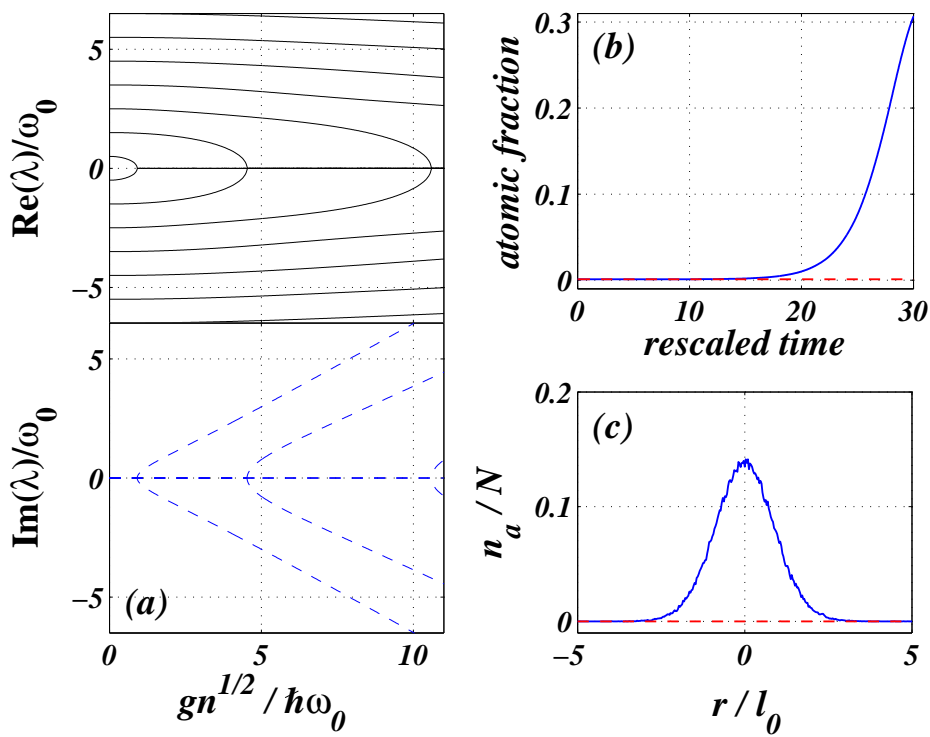}
\caption{Confinement effects on homonuclear dissociation: (a) Real and imaginary parts of the characteristic dissociation frequencies as a function of atom-molecule coupling strength, for an harmonic trap with frequency $\omz$. The coupling is resonant $\delta=0$, and the molecular condensate is noninteracting, $U_m=0$. Complex frequencies appear for $g\sqrt{\nav}>\hbar\omega_0$ (b) Total atomic fraction obtained in a mean-field dynamical calculation, triggered by Gaussian zero-average noise, as a function of rescaled time for $g\sqrt{\nav}/\hbar\omz=0.9$ (dashed red line) and for $g\sqrt{\nav}/\hbar\omz=1.0$ (solid blue line), with $\delta=U_m=0$ (c) Atomic density profiles from the same dynamical simulation at $\tau=30$. Amplification is clearly attained above the expected threshold.}
\label{fig1}
\end{figure}

In Fig. \ref{fig1}a we plot the characteristic frequencies of Eqs. (\ref{stabo})-(\ref{stabt}) in 1D with $\delta=U_m=0$ and a finite-size undepleted pump $\phi$ obtained from Eq. (\ref{mgp}), by expanding 
$\phi(\bfr)=\sum_j \varphi_j \chi_j(\bfr)$, $\hPsi(\bfr,t)=\sum_j \ha_j(t)\chi_j(\bfr)$ in the basis of harmonic-oscillator solutions $\chi_j(\bfr)$, and diagonalizing the resulting linear set of equations for the dynamics of $\ha_j(t)$ and $\ha_j^\dag(t)$,
\begin{equation}
i\frac{d}{dt}\left(\begin{array}{c}
{\bf {\hat a}}\\
{\bf {\hat a}^\dag}
\end{array}\right)=
\left(\begin{array}{cc}
\Omega&\Phi\\
-\Phi^\dag&-\Omega
\end{array}\right)
\left(\begin{array}{c}
{\bf {\hat a}}\\
{\bf {\hat a}^\dag}
\end{array}\right),
\end{equation}
with ${\bf {\hat a}}=(\ha_0,\ha_1,...,\ha_j,...)$, ${\bf {\hat a}^\dag}=(\ha^\dag_0,\ha^\dag_1,...,\ha^\dag_j,...)$, $\Omega_{ij}=\omz\delta_{ij} (j+1/2)$, $\Phi_{ij}=(g/\hbar)\sum_l \varphi_l\langle\chi_j\chi_l|\chi_i\rangle$. Identical eigenfrequency plots were obtained from a discrete Fourier transform diagonalization. Looking at the eigenfrequencies dependence on the atom-molecule conversion frequency $g\sqrt{\nav}/\hbar$, it is evident that complex frequencies, corresponding to an initially exponential molecular gain, are only obtained if the coupling frequency is larger than the trap frequency. Thus, for $g\sqrt{\nav}/\hbar\omz<\beta$ where $\beta$ is a geometrical factor of order one, the molecular condensate is stabilized by atomic zero point motion. The ensuing dynamics, including molecular BEC depletion, is depicted in Fig. \ref{fig1}b where atomic fraction is plotted as a function of time for $g\sqrt{\nav}$ just below the predicted amplification threshold (dashed red line) of $g_c\approx 0.91\hbar\omega/\sqrt{\nav}$ and slightly above it (solid blue line). The equations were solved in a mean-field approach, simulating initial spontaneous dissociation by a complex Gaussian white noise with zero mean, corresponding to atomic quantum fluctuations in the truncated Wigner representation \cite{Santagiustina98,Gatti97}. This technique is not as complete as full stochastic calculations \cite{Kheruntsyan02,Kheruntsyan05}, but it is sufficient to accurately predict the onset of instability. For $g<g_c$ all frequencies are real and there is no observed dynamical gain, whereas for $g$ just above the critical value, molecular amplification takes place. The corresponding atomic density profiles at time $\omz t=30$ are plotted in Fig. \ref{fig1}c. 
 
\section{Dissociation of a heteronuclear molecular BEC}
\label{heteronuc}

Having established that molecular BEC dissociation depends critically on trap parameters and that the stimulated process can be arrested altogether by sufficiently tight confinement, we turn to the case of heteronuclear diatomic molecules. The main difference from the homonuclear case is that due to differences in mass and dynamical polarizability, the dissociation fragments see effectively different trapping potentials and have different healing lengths. Thus we need to reformulate the amplification conditions accordingly.

We consider the dissociation reaction $AB\rightarrow A+B$, where diatomic $AB$ molecules are fragmented into their $A$ and $B$ constituent atoms, in an off-resonant optical trap. We denote the atomic masses by $m_A$, $m_B$ and the atomic dynamical polarizabilities by $\alpha_A$ and $\alpha_B$, respectively. The atomic trap frequencies experienced by the atoms are proportional to the square root of the polarizability to mass ratio,
\begin{equation}
\omega_{A,B}^2\propto\frac{\alpha_{A,B}}{m_{A,B}}~.
\end{equation} 
Since the polarizability of the loosely bound $AB$ molecules is approximately $\alpha_A+\alpha_B$, and their mass is $m_A+m_B$, the molecular trapping potential will be given to a good accuracy by,
\begin{equation}
\omega_{AB}^2=\frac{m_A\omega_A^2+m_B\omega_B^2}{m_A+m_B},
\end{equation}
reducing as required, to the atomic trap frequency when $\omega_A=\omega_B=\omega_0$.

\begin{figure}
\centering
\includegraphics[scale=0.7,angle=0]{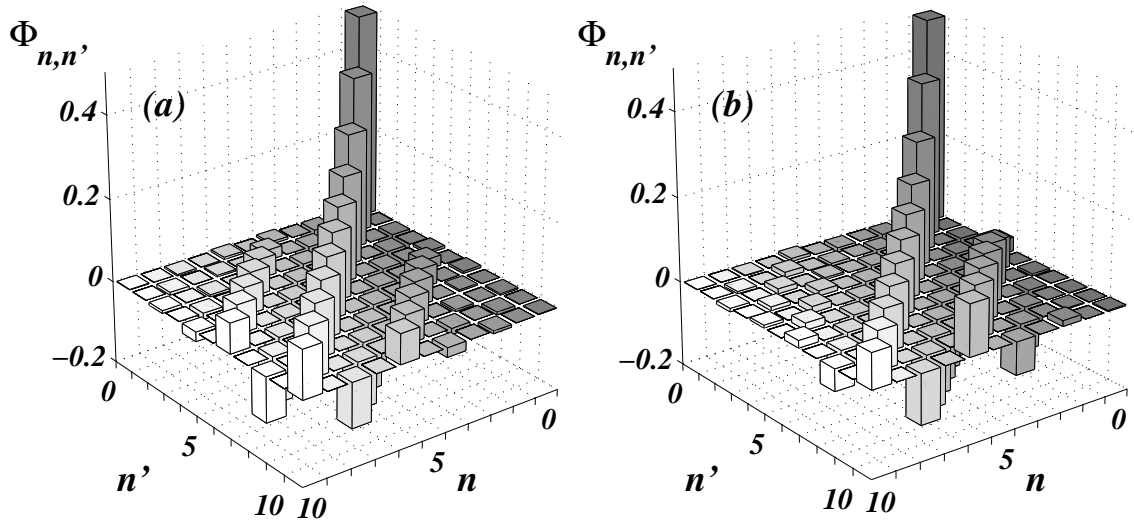}
\caption{Elements of the coupling matrix $\Phi_{nn'}$ with $U_m\nav/\hbar\omega_{AB}=10$, for $m_A=m_B$ (a) and for $m_B=2m_A$ (b). The even parity of the molecular condensate wavefunction forbids coupling between trap states with opposite parity. Diagonal elements dominate the coupling matrix.}
\label{fig2}
\end{figure}

Given these trap frequencies, the dynamical equations for the atomic field operators $\hPsi_{A,B}$, linearized about the molecular order parameter $\phi(\bfr)e^{-i\mu t/\hbar}$ obtained from the stationary GP equation,
\begin{equation}
\left[-\frac{\hbar^2}{2(m_A+m_B)}\nabla^2+\frac{(m_A+m_B)\omega_{AB}^2}{2}r^2+U_m|\phi|^2\right]\hphi=\mu\phi~,
\label{gpab}
\end{equation}
assume the form
\begin{eqnarray}
\label{abo}
i\hbar\frac{\partial}{\partial t}\hPsi_A&=&\left[-\frac{\hbar^2}{2m_A}\nabla^2+\delta+\frac{m_A\omega_A^2}{2}r^2\right]\hPsi_A+\frac{g}{2}\phi\hPsi_B^\dag,\\
\label{abt}
-i\hbar\frac{\partial}{\partial t}\hPsi_B^\dag&=&\left[-\frac{\hbar^2}{2m_B}\nabla^2+\delta+\frac{m_B\omega_B^2}{2}r^2\right]\hPsi_B^\dag+\frac{g}{2}\phi^*\hPsi_A~.
\end{eqnarray}
In what follows we shall assume $\delta=0$. Rescaling length as $\bfr\rightarrow\bfr/\lm$ with $\lm=\sqrt{\hbar/[(m_A+m_B)\omega_{AB}]}$, normalizing $\Phi=\phi/\sqrt{N}$ (so that $\int |\Phi(\bfr)|^2d\bfr=1$), and defining the dimensionless time $\tau=\omega_{AB} t$, Eqs.(\ref{gpab})-(\ref{abt}) transform to
\begin{equation}
\left[-\frac{1}{2}\nabla^2+\frac{r^2}{2}+\frac{\um\nav}{\hbar\omega_{AB}}|\Phi|^2\right]\Phi=\frac{\mu}{\hbar\omega_{AB}}\Phi
\label{gpabdl}
\end{equation}
\begin{eqnarray}
\label{abodl}
i\frac{\partial}{\partial \tau}\hPsi_A&=&\left[-\frac{M}{2m_A}\nabla^2+\frac{1}{2}\frac{m_A}{M}\left(\frac{\omega_A}{\omega_{AB}}\right)^2r^2\right]\hPsi_A+\frac{g\sqrt{\nav}}{2\hbar\omega_{AB}}\Phi\hPsi_B^\dag,\\
\label{abtdl}
i\frac{\partial}{\partial \tau}\hPsi_B^\dag&=&\left[\frac{M}{2m_B}\nabla^2-\frac{1}{2}\frac{m_B}{M}\left(\frac{\omega_B}{\omega_{AB}}\right)^2r^2\right]\hPsi_B-\frac{g\sqrt{\nav}}{2\hbar\omega_{AB}}\Phi^*\hPsi_A,
\end{eqnarray}
where $M\equiv m_A+m_B$ is the molecular mass and $\nav\equiv N/\lm^d$ is the average molecular density with $d$ denoting the number of dimensions. The characteristic frequencies of the set (\ref{abodl})-(\ref{abtdl}) are obtained by expanding the atomic field operators in terms of eigenmodes of the respective atomic traps, e.g. in 1D,
\begin{equation}
\hPsi_{A}=\sum_n \ha_n\chi_n^{A}(x)~~,~~ \hPsi_{B}=\sum_n \hb_n\chi_n^{B}(x)
\end{equation}
with
\begin{equation}
\chi_n^{A,B}(x)={\cal N}H_n\left(\sqrt{\frac{m_{A,B}\omega_{A,B}}{M\omega_{AB}}}x\right)\exp\left(-\frac{m_{A,B}\omega_{A,B}}{2M\omega_{AB}}x^2\right)~,
\end{equation}
where ${\cal N}=\frac{1}{\pi^{1/4}\sqrt{2^n n !}}\left(\frac{m_{A,B}\omega_{A,B}}{M\omega_{AB}}\right)^{1/4}$ and $H_n(z)=(-1)^ne^{z^2}\frac{d^n}{dz^n}e^{-z^2}$. This transformation reduces the dynamical equations into a coupled set of equations for the mode annihilation operators $\ha_n$ and $\hb_n$,
 \begin{eqnarray}
i\dot\ha_n&=&\left(n+\frac{1}{2}\right)\frac{\omega_A}{\omega_{AB}}\ha_n+\frac{g\sqrt{\nav}}{2\hbar\omega_{AB}}\sum_{n'}\Phi_{nn'}\hb^\dag_{n'}\\
i\dot\hb^\dag_n&=&-\left(n+\frac{1}{2}\right)\frac{\omega_B}{\omega_{AB}}\hb^\dag_n-\frac{g\sqrt{\nav}}{2\hbar\omega_{AB}}\sum_{n'}\Phi_{n'n}^*\ha_{n'}
\end{eqnarray}
where $\Phi_{nn'}=\int dx \chi^A_n(x)\Phi(x)\chi^B_{n'}(x)$. The coupling matrix elements $\Phi_{nn'}$ are plotted in Fig. \ref{fig2} for $1\le n,n'\le10$. Since the molecular condensate wavefunction possesses even parity, only trap modes with equal parity are coupled by the Feshbach interaction term. Moreover, because the molecular density is close to the ground state of the atomic trap, the diagonal elements dominate the coupling matrix. Neglecting all off-diagonal terms (this resembles the coupling of the the modes $\bfq$ and $-\bfq$ in the uniform case), we arrive at the characteristic frequencies,
\begin{equation}
\hbar\lambda_{\pm n}=\frac{1}{2}\left[\epsilon_n^A-\epsilon_n^B\pm\sqrt{(\epsilon_n^A+\epsilon_n^B)^2-\nav g^2\Phi_{nn}^2}\right],~n=0,1,2,...
\label{heteroeigen}
\end{equation}
where $\epsilon_n^{A,B}=\hbar\omega_{A,B}(n+1/2)$ are the respective harmonic oscillator energy levels. Thus, if the effect of off diagonal $\Phi_{n'n}$ elements can be neglected, amplification  will be attained only if $\sqrt{\nav}g\Phi_{nn}>\hbar(\omega_A+\omega_B)(n+1/2)$ for some $n$. Since $\Phi_{nn}<1$ for all $n$, we obtain the amplification condition,
\begin{equation}
|\sqrt{\nav}g\Phi_{00}|>\frac{\hbar(\omega_A+\omega_B)}{2}~.
\label{heteroampcond}
\end{equation} 

\begin{figure}
\centering
\includegraphics[scale=0.7,angle=0]{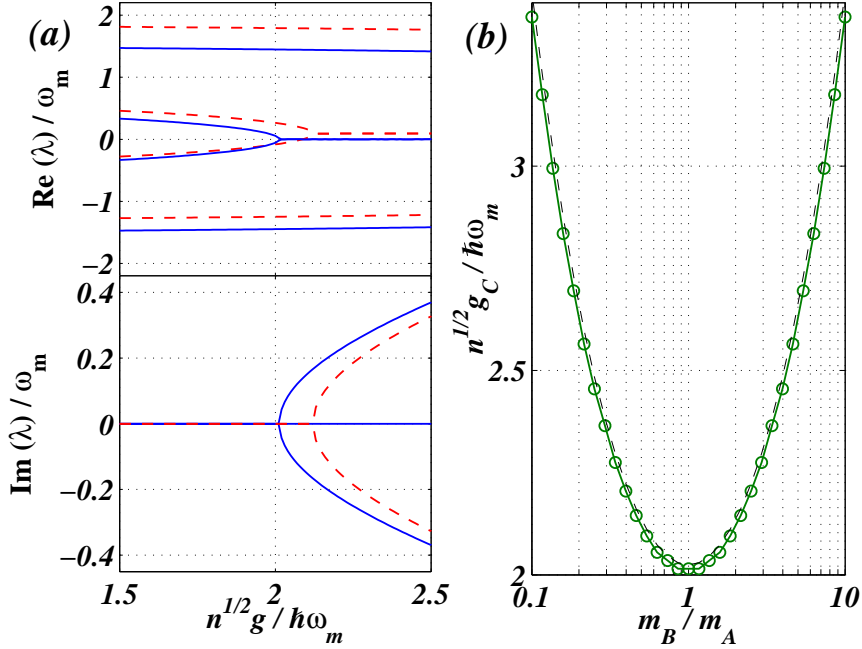}
\caption{Confinement effects on heteronuclear dissociation: (a) Real and imaginary parts of the first four characteristic dissociation frequencies, with $U_m\nav/\hbar\omega_{AB}=10$, for $m_A/\alpha_A=m_B/\alpha_B$ (solid blue line) and for $m_A/\alpha_A=m_B/2\alpha_B$. The critical value for stability to dissociation depends on the ratio $m_B\alpha_A/m_A\alpha_B$. (b) Dependence of critical coupling strength on mass ratio $m_B/m_A$ for $U_m\nav/\hbar\omega_{AB}=10$ and $\alpha_A=\alpha_B$. The exact value obtained by direct diagonalization (solid green line, circles) is compared with the approximate critical coupling from Eq. (\ref{heteroampcond}) (dashed black line).}
\label{fig3}
\end{figure}

In Fig. \ref{fig3}a we plot the first four characteristic frequency obtained by {\it full diagonalization} including all off diagonal $\Phi_{nn'}$ terms, as a function of the coupling strength $g\sqrt{n}$, for $m_A=m_B$ (solid blue line) and for $m_B=2m_A$ (red dashed line). The optical polarizabilities were assumed for simplicity, to be identical $\alpha_A=\alpha_B$. The resulting eigenvalues turn out to be almost identical to the approximate formula (\ref{heteroeigen}). For equal masses, we have $\omega_A=\omega_B=\omega_{AB}$, so that amplification is attained for $g\sqrt{\nav}>\hbar\omega_{AB}/\Phi_{00}=2\hbar\omega_{AB}$. We note that the factor two difference from the homonuclear case results in from the distinguishability of the dissociation fragments. A similar factor of two separates the nondegenerate parametric amplifier from its degenerate counterpart. For $m_B=2m_A$ and $\alpha_A=\alpha_B$ we have $\omega_A=\sqrt{2}\omega_B=\sqrt{3/2}\omega_{AB}$, so that $\omega_A+\omega_B=\frac{\sqrt{3}(1+\sqrt{2})}{2}\omega_{AB}\approx 2.1\omega_{AB}$. Consequently, complex frequencies begin to appear for $g\sqrt{\nav}\approx 2.1\hbar\omega_{AB}$, as shown in Fig. \ref{fig3}a. In Fig \ref{fig3}b the critical coupling strength for stimulated dissociation, is plotted as a function of the mass ratio $m_B/m_A$. The solid green line is obtained by direct diagonalization, whereas the dashed black line corresponds to the approximate amplification condition (\ref{heteroampcond}), evidently giving an excellent estimate of the amplification threshold. Numerical simulations of dynamics, carried out just below and just above the anticipated critical coupling, clearly show the transition from small oscillations and an initial exponential gain. Thus the outcome of dissociation will depend on the mass ratio between the dissociation fragments.

\section{Control of two-channel dissociation by confinement}
\label{multi}

\begin{figure}
\centering
\includegraphics[scale=0.7,angle=0]{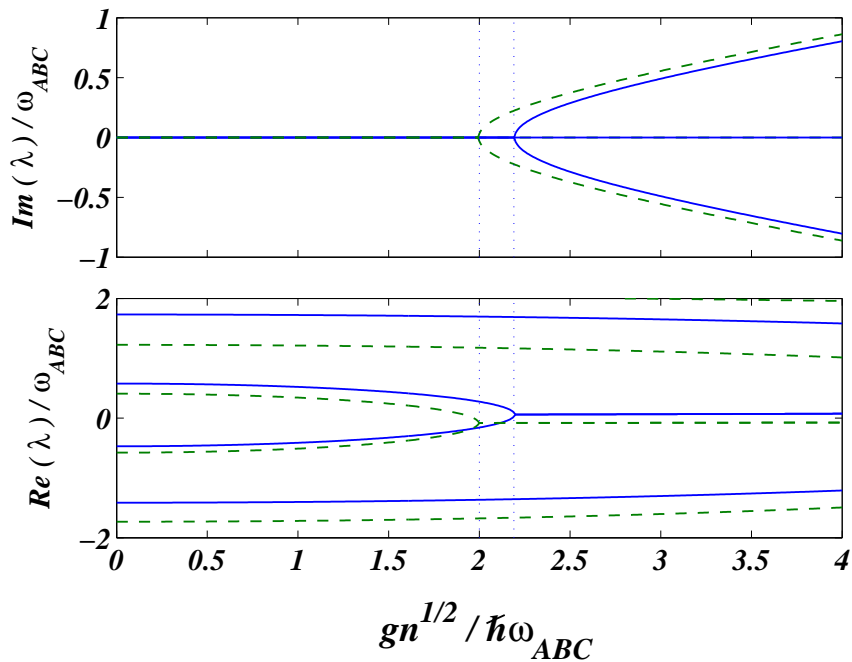}
\caption{Real and imaginary parts of the first eight characteristic dissociation frequencies for two-channel dissociation with $m_A/\alpha_A=m_B/\alpha_B=m_C/2\alpha_C$ and $U_m\nav/\hbar\omega_{AB}=10$, as a function of the coupling parameter $g\sqrt{\nav}/\hbar\omega_{ABC}$. Solid blue lines denote the frequencies associated with $A+BC$ channel whereas dashed green lines correspond to the $AB+C$ channel. Vertical dotted lines mark the critical coupling strength of each channel. In between these lines, amplification will be attained only in the $AB+C$ route.}
\label{fig4}
\end{figure}

The possibility to control multichannel dissociation through confinement is a straightforward extension of the previous sections. Consider a quantum gas of triatomic $ABC$ molecules which may undergo dissociation into either $A+BC$ or into $AB+C$ \cite{Moore02}. This is a second-quantized version of the familiar coherent control paradigm.  Given that the constituent atoms are subject to trap frequencies $\omega_{A,B,C}$ which are proportional to $\sqrt{\alpha_{A,B,C}/m_{A,B,C}}$ respectively, the trap frequencies for the $BC$ and $AB$ fragments can be approximately given as $\omega_{BC}^2=\frac{m_B\omega_B^2+m_C\omega_C^2}{m_B+m_C}$ and $\omega_{AB}^2=\frac{m_A\omega_A^2+m_B\omega_B^2}{m_A+m_B}$ , respectively. The trap frequency for $ABC$ molecules is  $\omega_{ABC}^2=\frac{m_A\omega_A^2+m_B\omega_B^2+m_C\omega_C^2}{m_A+m_B+m_C}$. Mass differences between the constituent atoms translate to differences between the $m_A/(m_B+m_C)$ and the $m_C/(m_A+m_B)$ fragment mass ratios, resulting in different amplification thresholds. Consequently it is possible to control the dissociation by variation of the trapping field.

Rescaling as before $\bfr\rightarrow\bfr/\labc$ with $\labc=\sqrt{\hbar/(M\omega_{ABC})}$, $t\rightarrow\tau=\omega_{ABC} t$, $\phi\rightarrow\Phi=\phi/\sqrt{N}$, the two-channel linearized equations (assuming $\Delta_1=\Delta_2=\mu/2$ with $\Delta_{1,2}$ being the respective $A+BC$ and $AB+C$ channel detuning), are a simple extension of Eqs. (\ref{gpabdl})-(\ref{abtdl}),
\begin{equation}
\left[-\frac{1}{2}\nabla^2+\frac{r^2}{2}+\frac{\um\nav}{\hbar\omega_{ABC}}|\Phi|^2\right]\Phi=\frac{\mu}{\hbar\omega_{ABC}}\Phi
\label{gpabcdl}
\end{equation}
\begin{eqnarray}
\label{abcodl}
i\frac{\partial}{\partial \tau}\hPsi_A&=&\left[-\frac{M}{2m_A}\nabla^2+\frac{1}{2}\frac{m_A}{M}\left(\frac{\omega_A}{\omega_{ABC}}\right)^2r^2\right]\hPsi_A+\frac{g_1\sqrt{\nav}}{2\hbar\omega_{ABC}}\Phi\hPsi_{BC}^\dag,\\
\label{abctdl}
i\frac{\partial}{\partial \tau}\hPsi_{BC}^\dag&=&\left[\frac{M}{2m_{BC}}\nabla^2-\frac{1}{2}\frac{m_{BC}}{M}\left(\frac{\omega_{BC}}{\omega_{ABC}}\right)^2r^2\right]\hPsi_{BC}^\dag-\frac{g_1\sqrt{\nav}}{2\hbar\omega_{ABC}}\Phi^*\hPsi_A,\\
\label{abcthdl}
i\frac{\partial}{\partial \tau}\hPsi_C&=&\left[-\frac{M}{2m_C}\nabla^2+\frac{1}{2}\frac{m_C}{M}\left(\frac{\omega_C}{\omega_{ABC}}\right)^2r^2\right]\hPsi_C+\frac{g_2\sqrt{\nav}}{2\hbar\omega_{ABC}}\Phi\hPsi_{AB}^\dag,\\
\label{abcfdl}
i\frac{\partial}{\partial \tau}\hPsi_{AB}^\dag&=&\left[\frac{M}{2m_{AB}}\nabla^2-\frac{1}{2}\frac{m_{AB}}{M}\left(\frac{\omega_{AB}}{\omega_{ABC}}\right)^2r^2\right]\hPsi_{AB}^\dag-\frac{g_2\sqrt{\nav}}{2\hbar\omega_{ABC}}\Phi^*\hPsi_C.
\end{eqnarray}
Here $g1$ and $g_2$ denote the atom molecule coupling strength in the $A+BC$ and $AB+C$ channel, respectively.Characteristic frequencies are again found by expanding all the field operators of the fragments, in terms of the respective trap modes and diagonalizing the resulting set of equations for the mode annihilation and creation operators. Since the two channels are only coupled through the mutual pump field, the eigenvalue problems separate and the approximate eigenfrequencies are the same as in Eq. (\ref{heteroeigen}),
\begin{eqnarray}
\hbar\lambda_{\pm n}^{A+BC}&\approx&\frac{1}{2}\left[\epsilon_n^A-\epsilon_n^{BC}\pm\sqrt{(\epsilon_n^A+\epsilon_n^{BC})^2-\nav g_1^2\Phi_{1,nn}^2}\right]\nonumber\\
\hbar\lambda_{\pm n}^{AB+C}&\approx&\frac{1}{2}\left[\epsilon_n^C-\epsilon_n^{AB}\pm\sqrt{(\epsilon_n^C+\epsilon_n^{AB})^2-\nav g_2^2\Phi_{2,nn}^2}\right],
\label{tceigen}
\end{eqnarray}
where $\Phi_{1,nn'}=\int dx \chi^A_n(x)\Phi(x)\chi^{BC}_{n'}(x)$, $\Phi_{2,nn'}=\int dx \chi^{AB}_n(x)\Phi(x)\chi^C_{n'}(x)$ and $\chi_n^\alpha(x)$ are the respective oscillator wavefunctions for species $\alpha$. Consequently, the amplification condition for each channel is roughly,
\begin{equation}
\sqrt{\nav}|g_1\Phi_{1,00}|>\frac{\hbar(\omega_A+\omega_{BC})}{2}~,~
\sqrt{\nav}|g_2\Phi_{2,00}|>\frac{\hbar(\omega_C+\omega_{AB})}{2}.
\label{tmampcond}
\end{equation}

\begin{figure}
\centering
\includegraphics[scale=0.7,angle=0]{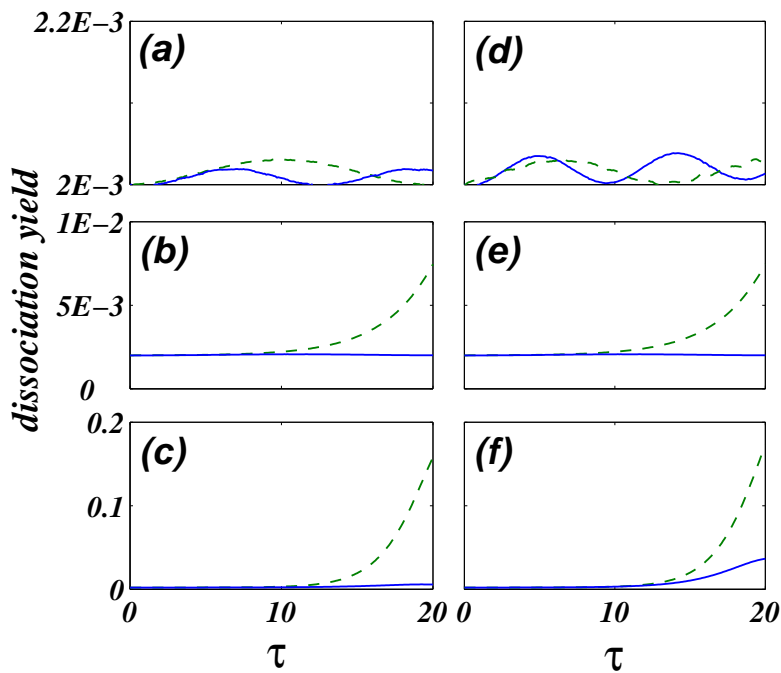}
\caption{Fraction of dissociated atoms in the $A+BC$ channel (solid blue line) and the $AB+C$ channel (dashed green line) as a function of rescaled time $\tau=\omega_{ABC}t$. Atomic masses, optical polarizabilities, and the molecular interaction strength are the same as in Fig. \ref{fig4}. Coupling strengths are: (a) $g\sqrt{\nav}/\hbar\omega_{ABC}=1.9$, (b) $g\sqrt{\nav}/\hbar\omega_{ABC}=2.1$, (c) $g\sqrt{\nav}/\hbar\omega_{ABC}=2.3$. In (d)-(f) we retain the coupling strength, length and time units of (b) and multiply all trap frequencies by a 'widening factor' $w$ of 0.8 (d), 1.0 (e), and 1.2 (f). The channel population distribution depends on trap size.}
\label{fig5}
\end{figure}

\begin{figure}
\centering
\includegraphics[scale=0.7,angle=0]{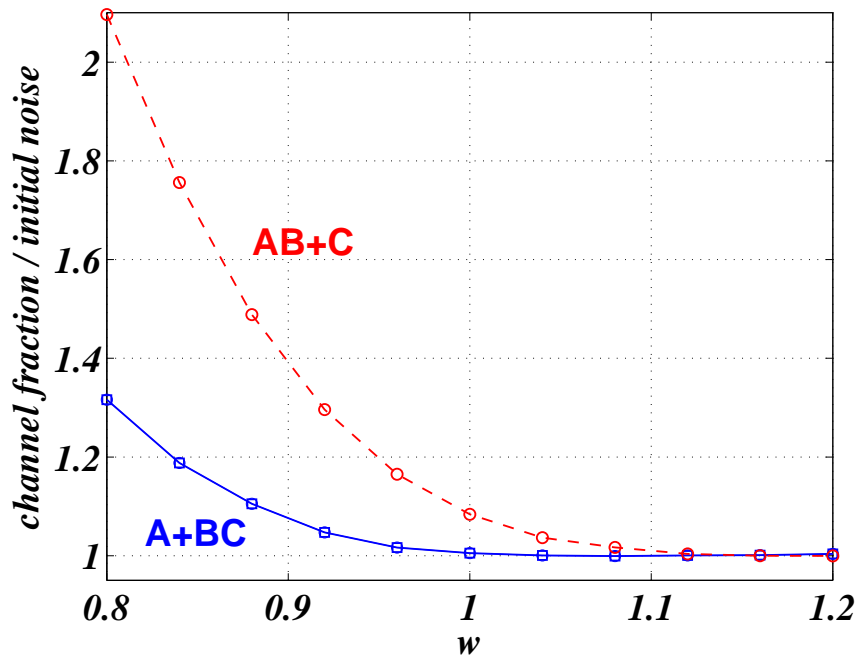}
\caption{Channel populations at $\tau=15$ as a function of the widening parameter $w$. The dimensionless time is given in terms of the $w=1$ trap period. All parameters are the same as in Figs. \ref{fig5}d-\ref{fig5}f.}
\label{fig6}
\end{figure}

By judicious choice of the confining potential it is thus possible to find regimes where gain is attained only in one of the two dissociation channels, even when $g_1=g_2$ and both channels are equally coupled to the molecular BEC. As an example, we shall consider an arbitrary case where $m_A/\alpha_A=m_B/\alpha_B=m_C/2\alpha_C$, $g_1=g_2=g$, and $U_m\nav/\hbar\omega_{ABC}=10$, corresponding to $\mu=3.11\hbar\omega_{ABC}$. In Fig. \ref{fig4}, the exact characteristic frequencies $\hbar\lambda_{\pm n}^{A+BC}$ (solid blue line) and $\lambda_{\pm n}^{AB+C}$ (dashed green line) are plotted as a function of the coupling strength. The critical values for amplification are $g^{(c)}_{A+BC}\sqrt{\nav}/\hbar\omega_{ABC}=2.1960$ and $g^{(c)}_{AB+C}\sqrt{\nav}/\hbar\omega_{ABC}=1.9960$, in very good agreement with the approximate conditions (\ref{tmampcond}). In between these critical coupling values, there lies a window where exponential gain will be attained only in the $AB+C$ channel. This prediction is supported by the numerical simulations of dynamics (including pump depletion) for various values of $g\sqrt{\nav}/\hbar\omega_{ABC}$, shown in Figs. \ref{fig5}a through  \ref{fig5}c. When $g\sqrt{\nav}/\hbar\omega_{ABC}=1.9$ (Fig. \ref{fig5}a) the coupling is subcritical and both channels are stabilized against dissociation. At $g\sqrt{\nav}/\hbar\omega_{ABC}=2.1$ (Fig. \ref{fig5}b) only the $AB+C$ channel is amplified, and for supercritical $g\sqrt{\nav}/\hbar\omega_{ABC}=2.3$ (Fig. \ref{fig5}c) atoms are produced in both channels. Control can be attained either by varying the coupling strength $g$ or by changing the trap widths. If the trap is tightened or loosened so that all trap frequencies are multiplied by a factor $w$ (e.g. by variation of the trap field strength) while keeping fixed $g$, $U_m$ and the total number of particles $N$, the critical coupling-strength values $g^{(c)}_{A+BC}$ and $g^{(c)}_{AB+C}$ get multiplied roughly by a factor of $w^{3/4}$, because trap level spacing is multiplied by $w$ and the average density is multiplied by $w^{1/4}$ (the exact factor in the presence of molecular interactions is slightly different, due to variation in the molecular condensate profile). For example, for $w=1.2$ we have $g^{(c)}(w=1.2)\approx 1.2^{3/4}\times g^{(c)}(w=1)=1.1465g^{(c)}(w=1)$. Retaining the dimensionless length and frequency scales of $w=1$, we obtain that $g^{(c)}_{A+BC}\approx 2.52$ and  $g^{(c)}_{AB+C}\approx2.29$. The actual values from numerical diagonalization turn out to be $g^{(c)}_{A+BC}=2.5000$ and $g^{(c)}_{AB+C}=2.2680$ indicating that the effect of molecular density-profile variation is negligible. Similarly, for $w=0.8$, the critical values are multiplied by $0.8^{3/4}=0.8459$ to give $g^{(c)}_{A+BC}\approx 1.86$ and  $g^{(c)}_{AB+C}\approx 1.69$ (in units of the $w=1$ trap frequency and length), compared to numerical values of $1.8760$ and $1.7080$, respectively. In Figs. \ref{fig5}d-\ref{fig5}f we plot a sequence of dynamical calculations keeping $g$, $U_m$ and $N$ fixed, and varying the trap frequency from $w=1.2$ (Fig \ref{fig5}d) through $w=1$  (Fig \ref{fig5}e), to $w=0.8$ (Fig \ref{fig5}f). This sequence corresponds to a realistic experimental scenario where only the trap frequency is varied and it clearly shows the dependence of the reaction outcome and the channel population ratio, upon the confining potential width. The atomic population at $\tau=15$ (in $w=1$ dimensionless frequency units) is plotted in Fig. \ref{fig6} as a function $w$, keeping $N$, $g$ and $U_m$ fixed. It is evident that the ratio between the channel populations can be controlled merely by variation of trap width.

\section{Experimental parameters}
\label{experiment}
Before closing, we assess the relevant experimental parameters required for the observation of the predicted supression of stimulated dissociation by tight confinement. For the majority of Feshbach-coupled diatomic systems the characteristic atom-molecule conversion frequency is much larger than the trap frequency, suggesting they are well-within the strong-coupling, nonlinear domain where stimulated gain should be observed. One example is the $^{23}$Na Feshbach resonance at 907 G \cite{Inouye98,Stenger99}, where the coupling strength is given by $g=\sqrt{\kappa U_a}$ \cite{Timmermans99,Holland01} with $U_a=4\pi\hbar^2a_s/m$, $\kappa/(2\pi\hbar)\approx4.6$ MHz, and $a_s\approx60$ Bohr. These values give $g/2\pi\hbar\approx 9\times10^{-3}$ Hz cm$^{3/2}$. Given a density of $n=10^{15}$ cm$^{-3}$, we have $g\sqrt{n}/2\pi\hbar\approx 284$ kHz, far above contemporary trap frequencies (we note that comparison of this Feshbach frequency to the atomic interaction frequency $U_a n/2\pi\hbar\approx 17.5$ kHz, justifies the neglect of atom-atom interactions in our calculations, even for substantial atomic populations and high densities. For $n=10^{13}$cm$^{-3}$, $g\sqrt{n}/2\pi\hbar$ drops to $28.4$ kHz, whereas $U_a n/2\pi\hbar$ is only $175$ Hz). For $\omz/2\pi=200 Hz$ we obtain that $N_c\approx 2\times10^{-2}$, so that stimulated dissociation will be  observed for any number of molecules in the condensate. It may be that a sufficiently weak Feshbach resonance can be found amongst the many resonances of $^{87}$Rb \cite{Marte02}.  

Collective dissociation thresholds, can however be observed in optical stimulated-Raman experimental setups \cite{Wynar00,Rom04,Winkler05,Ryu06,Stoferle06}. In this type of atom-molecule coupling, $g=\Omega_p\Omega_s/2\Delta_{2p}$ is an effective two-photon Rabi frequency. The one-photon Rabi frequencies $\Omega_{s,p}$ for the pump (inducing a bound-bound transition to an intermediate molecular state) and the Stokes (dissociating the intermediate state via a bound-free transition) lasers, respectively, are products of laser intensities by overlap integrals consisting of electronic transition dipole moments and Franck-Condon factors, and $\Delta_{2p}$ is the two-photon detuning from the intermediate bound state. Thus, the effective coupling strength can be controlled with great precision \cite{Rom04} through the adjustment of laser parameters. Most appealing are systems of molecules in optical lattices \cite{Jaksch02,Rom04,Ryu06,Stoferle06}, providing optical coupling to deeply bound internal molecular states, as well as tight trap frequencies of $\omz/2\pi=10-100$ kHz. Currently, these experiments seek to avoid collective effects in association by operating in the Mott-insulator regime with unit occupation numbers. However, provided that molecular condensates containing roughly $10^2-10^3$ particles per site be formed, their dissociation can demonstrate the expected amplification threshold. 

For $10^3$ $^{87}$Rb$_2$ molecules in a $1$ kHz trap, the critical value $g_c=\hbar^2/(m\sqrt{N\lz/\Lambda_3})$ of the interaction strength turns out to be approximately $2\pi\hbar\times 2.2\times10^{-5}$ Hz cm$^{3/2}$. The corresponding average atomic density is of order $10^{13}-10^{14}$ cm$^{-3}$ and $a_s\approx100$ bohr for $^{87}$Rb. Therefore, $Un\approx 75-750$ Hz for a fully dissociated gas, and atomic interactions are initially negligible with respect to the trap and coupling frequencies.One possible experiment will involve the preparation of a molecular BEC in a trap smaller than the resonance healing length, thus arresting dissociation even in the presence of coupling. A sudden change either in $g$ or in $\omz$ (which in a lattice can be achieved via switching the lattice wavelength or depth) will trigger the stimulated dissociation of the BEC \cite{Tikhonenkov07}.

While similar values for the critical parameters apply to the case of heteronuclear diatomic molecules, it is more difficult to estimate the pertinent experimental parameters for the heteronuclear triatomics described in section \ref{multi}, due to the lack of experimental or theoretical values for the required Feshbach resonance strengths.  

\section{Conclusions}
\label{conclusion}

The realization of dilute Bose condensed gases has produced ensembles with unprecedented mesoscopic coherence lengths of the order of several $\mu m$. These healing lengths are comparable with characteristic confining potential lengths, suggesting a wealth of interesting phenomena, such as the stabilization of condensates of attractively interacting atoms \cite{Ruprecht95,Bradley} and the generation of matter-wave solitons. The production of molecular BECs \cite{Regal03,Jochim03,Zwierlein03} opens the way for the utilization of this macroscopic coherence to generate a novel type of collective 'superchemistry', in which atoms are not treated individually but as one mesoscopic, coherent entity \cite{Heinzen00,Vardi02,Moore02}. Previous work has shown that bose stimulation effects should dominate atom-molecule dynamics in degenerate quantum bose gases. It was found that the resulting exponential gain produces extremely heightened selectivity in multichannel processes \cite{Moore02} and can lead to pattern formation in molecular dissociation \cite{Vardi02}.

Atom-molecule coupling is characterized by a resonance healing length. If confining trap sizes become comparable with this length, the molecular BEC can be stabilized against dissociation in the same way an attractively interacting BEC is stabilized against collapse \cite{Tikhonenkov07}. In this work we have used this stabilization effect to show that confinement can also be used to control multichannel dissociation. Differences in mass and polarizability of the dissociation fragments produce differences in the resonance healing lengths and effective trap lengths of the various rearrangement channels, resulting in different amplification thresholds, even when the coupling strengths are equal. Consequently, the channel population ratio depends critically on the trap width and geometry. This effect highlights another novel feature of superchemistry in that reaction yields depend on reaction vessels.

\end{document}